\documentclass[aps,prl,reprint]{revtex4-1}
\usepackage{graphicx}

\begin{document}
\title{Nitric Oxide as stress inducer and synchronizer of $p53$ dynamics}
\author{Gurumayum Reenaroy Devi$^1$}
\author{Md. Jahoor Alam$^2$}
\author{R.K. Brojen Singh$^{2}$}
\email{brojen@jnu.ac.in}
\affiliation{$^1$Centre for Interdisciplinary Research in Basic Sciences, Jamia Millia Islamia, New Delhi-110025, India. \\
$^2$School of Computational and Integrative Sciences, Jawaharlal Nehru University, New Delhi-110067, India.}

\begin{abstract}
We study how the temporal behaviours of $p53$ and $MDM2$ are affected by stress inducing bioactive molecules $NO$ (Nitric Oxide) in the $p53-MDM2-NO$ regulatory network. We also study synchronization among a group of identical stress systems arranged in a three dimentional array with nearest neighbour diffusive coupling. The role of $NO$ and effect of noise are investigated. In the single system study, we have found three distinct types of temporal behaviour of $p53$, namely, oscillation death, damped oscillation and sustain oscillation, depending on the amount of stress induced by the $NO$ concentration, indicating how $p53$ responds to the incoming stress. The correlation among the coupled systems increases as the value of coupling constant ($\epsilon$) is increased ($\gamma$ increases) and becomes constant after certain value of $\epsilon$. The permutation entropy spectra $H(\epsilon)$ for $p53$ and $MDM2$ as a function of $\epsilon$ are found to be different due to direct and indirect interaction of $NO$ with the respective proteins. $\gamma$ versus $\epsilon$ for $p53$ and $MDM2$ are found to be similar in deterministic approach, but different in stochastic approach and the separation between $\gamma$ of the respective proteins as a function of $\epsilon$ decreases as system size increases. The role of $NO$ is found to be twofold: stress induced by it is prominent at small and large values of $\epsilon$ but synchrony inducing by it dominates in moderate range of $\epsilon$. Excess stress induce apoptosis to the system.
\end{abstract}

\maketitle

\section{Introduction}

The tumour suppressor p53 is a transcription factor which functions to protect cells from malignant transformation \cite{Vou}. It is believed that most of the tumours are developed after the loss of p53 function \cite{Lev,Vog,Lane}. It has been shown that that p53 participates in various regulating processes, for example, cell cycle arrest \cite{livi}, cell differentiation \cite{murr}, cell senescence \cite{serr}, DNA repair \cite{saka}, cell apoptosis \cite{yoni} etc. But when the cell experience stress, p53 induce both cell cycle arrest and apoptosis by activating the p53 responsive target genes corresponding to the amount of stress \cite{Bat}. In normal stress free cell, it exhibits short half life time and usually maintained at low level through ubiquitin-dependent proteolysis \cite{Kub,Cho}. Regulation of p53 protein is maintained by interaction with the MDM2 protein which is inhibitor of p53 \cite{Hau}. MDM2 is the antagonist of p53 and has two functions, first as a ubiquitin ligase for tumour suppressor p53 \cite{Hon} and secondly shuttles p53 from the nucleus to the cytoplasm, where degradation of p53 takes place \cite{Tao}. MDM2 itself is transcriptionally regulated by p53, thereby establishing a negative feedback loop in which increased levels of p53 activate increased expression of MDM2, which targets p53 for degradation \cite{Wu}.

Several cellular stresses including DNA damage, oncogene activation, hypoxia, cellular ribonucleotide depletion, mitotic spindle damage and nitric oxide (NO) production activate p53 \cite{She}. Upon exposure to stress, p53 is activated but can exist upto certain range of stress, and then stabilized predominantly by post-translational modification, such as phosphorylation events \cite{Mee}, acetylation \cite{Gu} or glycosylation \cite{Shaw}. Nitric oxide (NO) is an important bioactive and messenger molecule involved in a variety of physiological processes of living system \cite{Stern}. NO can be synthesized either by specific nitric-oxide synthases (NOS), which metabolize arginine to citrulline with the formation of NO or by non enzymatic mechanisms which reduces nitrite to NO \cite{Zwe}. It is a membrane permeable and extremely short-lived molecule about half-life of 1-10 s \cite{Xin}. Because of this permeability quality of NO and its widespread availability, it can also function as a signalling molecule for cellular communication \cite{torr}.

Cellular stress induce oscillations in the dynamics of p53 \cite{Gal,Bar} because stress affects p53 in two ways: firstly increasing p53 concentration in the cell and secondly enhances the transcription of genes regulated by DNA sequences where p53 binds \cite{San}. NO induces p53 accumulation and MDM2 degradation by phosphorylation\cite{Nic,Wan,Wang1}. Further, p53 phosphorylation initiates ubiquitin-dependent degradation of MDM2 and consequently regulates the stability of both p53 and MDM2 \cite{Sung}. However, single cell dynamics affected by NO needs to be investigated in wider sense because of the two-fold role of NO which acts as stress inducer and synchronizing agent that may lead the cell in various complex phenomena. Further, when a group of such cells are allowed to interact via NO, how do cells communicate via nitric oxide diffusion needed to be investigated in both theoretical and experimental situation. The present understanding of how cellular networks interact via direct and indirect interaction at molecular level still needs to be investigated extensively in order to capture complexities due to different levels of these interaction.

The inter and intra molecular interactions  and dynamics in heterogeneous, crowded molecular population and complex environmental fluctuations in a living cell is stochastic in nature \cite{yan,bres}. There have various approaches in the study of transport phenomena in finite system which show the possibility of retrieving macro level stochastic transport phenomenon in finite system from the dynamical studies in underlying $Hamilton's$ system \cite{yan1}. Further, radiation drive \cite{bo} and damage-drive random signal \cite{bo1} on p53 is incorporated in the system systematically with intrinsic noise which exhibit oscillation in the dynamics and argued that this stress conditions in p53-Mdm2 system drive the system to oscillation in p53 dynamics and noise in the stochastic dynamics dampen the p53 dynamics \cite{bo2,bo3}. It has also been studied noise induced oscillation in different systems, weak noise induced stochastic resonance \cite{zak}, noise induced synchronization of coupled systems \cite{zho}, noise induced phase transition and noise induced dynamical change \cite{lind}. It is also shown in delay p53 regulatory model that the possibilities of inducing different p53 oscillatory dynamics interplay by noise and delay parameter \cite{bo4}. Different approach to retrieve macroscopic dynamics of a system from the microscopic description of the Hamiltonian system and general fluctuation-dissipation theorem was introduced to understand the role of noise which is capable of driving the system \cite{yan2}.

The manifestations of sustained and damped oscillations, and their transition from one state to another in p53 dynamics induced by DNA damage is systematically incorporated in the system \cite{yan4,yan5} and was found to be sensitive to the noise and delay parameter in the system model \cite{bo4,sheng}. Delay induced oscillatory dynamics in gene expression \cite{wangq}, noise induced transition of oscillatory dynamics \cite{bo5} were also studied in the contex of stochastic systems. We address some of the interesting phenomena how do NO drives the system at various states, plays various roles in regulating the system and process signal among the interacting systems.

We organize our work as follows. We describe our model, methods to be applied in Material and Methods in section 2. The numerical results and discussions are presented in section 3 and conclusions are drawn based on the numerical results we obtained in section 4.

\begin{figure}
\centering{\includegraphics[width=100mm]{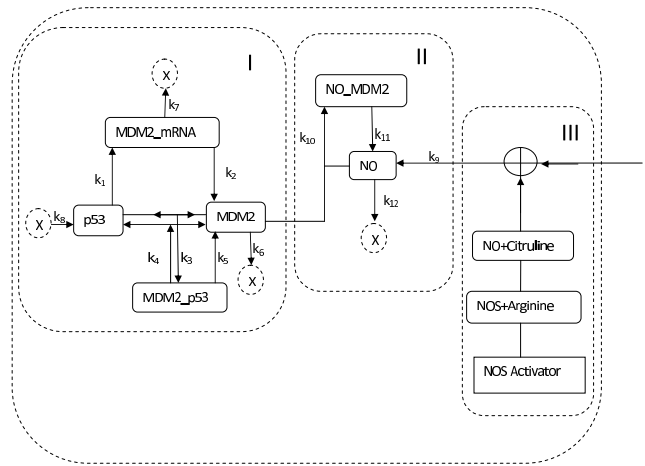}}
\caption{A schematic diagram of biochemical reaction network of p53-MDM2 Oscillator induced by NO.}
\end{figure}

\section{Materials and Methods}
\subsection{\bf NO induced p53-MDM2 Stress Model}

In this model, we extended p53-MDM2 regulatory network explained by Procter and Gray \cite{Pro} by allowing NO interaction with the network via MDM2 (Fig. 1). Feedback regulatory mechanism between $p53$ and $MDM2$ plays a key role in this network and can be described briefly as follows. $MDM2$ gene is transcriptionally activated by $p53$ to form $MDM2\_mRNA$ \cite{Gal,Pro} with rate $k_1$ which in turn synthesizes $MDM2$ protein in the system \cite{Pro,Fin} with rate $k_2$. $MDM2$ then interacts with $p53$ forming $MDM2\_p53$ complex \cite{Pro} with rate $k_3$, and this complex then dissociates to individual proteins again \cite{Mol} with rate $k_4$. This is followed by $p53$ degradation from dissociation of $MDM2\_p53$ with rate $k_5$ due to ubiquitination\cite{Hau,Kub,Moma}.  $MDM2$ and $MDM2\_mRNA$ degrate with $k_6$ and $k_7$ respectively and $p53$ is synthesized with rate $k_8$.
\begin{figure}
\centering{\includegraphics[width=100mm]{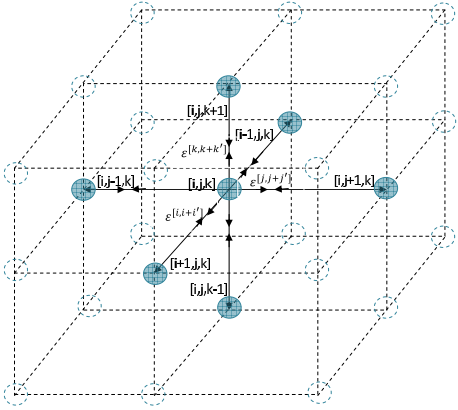}}
\caption{A schematic diagram of interacting systems in three dimentional network.}
\end{figure}
$NO$ is a small signalling, bioactive and diffusible molecule through cell membrane \cite{Stern} and is constantly synthesized inside cell through enzyme metabolism with rate $k_9$ \cite{Stern,Wood}. One of its important functions is to down regulate $MDM2$ level by forming a complex $NO\_MDM2$ \cite{Wan,Sci} with rate $k_{10}$ to upregulate $p53$ level causing normal to stress cell \cite {Stern}. After activation, $p53$ and $MDM2$ proteins remain localized in the nucleus to activate the target genes \cite{Chen,Lia}. This is followed by dissociation of the complex $NO\_MDM2$ with rate $k_{11}$ and degradation of $NO$ with rate $k_{12}$ \cite{Sung}. The molecular species notation and the reaction sets with their respective kinetic laws and rate constants are shown in Table 1 and Table 2.
\begin{figure}
\centering{\includegraphics[width=100mm]{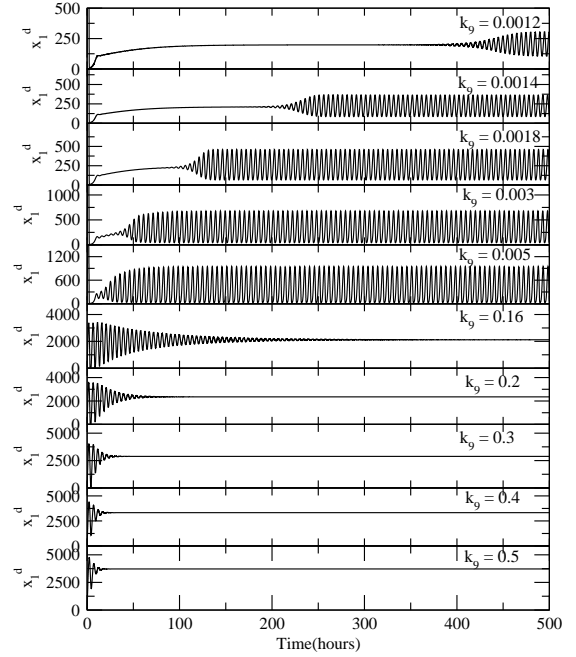}}
\caption{Temporal behaviors of p53 for different values of $k_9$ in Single cell by using Deterministic approach.}
\end{figure}

The biochemical network in Fig. 1 can be described deterministically by translating the reaction channels into a set of ordinary differential equations (ODE) using mass action law \cite{Mac}. If we denote the molecular species $p53$, $MDM2$, $MDM2\_p53$, $MDM2\_mRNA$, $NO$, $NO\_MDM2$ by $x_1$, $x_2$, $x_3$, $x_4$, $x_5$ and $x_6$ respectively also given in Table 1, then the molecular species vector for the model we consider is given by, ${\bf x}(t)==[x_1(t),x_2(t),x_3(t),x_4(t),x_5(t),x_6(t)]^T$, where T is the transpose of the vector, and the set of ODEs is given by,
\begin{eqnarray}
\label{det}
\frac{d}{dt}{\bf x}(t)={\bf M}(x_1,x_2,\dots,x_6;t)
\end{eqnarray}
where, ${\bf M}$ is a vector given by, ${\bf M}=[m_1,m_2,\dots,m_6]^T$: $m_1=-k_3x_1x_2+k_4x_3+k_8$, $m_2=k_2x_4-k_3x_1x_2+k_4x_3+k_5x_3-k_6x_2-k_{10}x_5x_2$, $m_3=k_3x_1x_2-k_4x_3-k_5x_3$, $m_4=k_1x_1-k_7x_4$, $m_5=k_9-k_{10}x_5x_2+k_{11}x_6-k_{12}x_5$, $m_6=k_{10}x_5x_2-k_{11}x_6$.

\section{Tables}
\begin{table}
\begin{center}
{\bf Table 1 List of Molecular Species} 
    \begin{tabular}{ | l | p{2cm} | p{3cm} | p{2cm} |}
       \hline \multicolumn{4}{}{} \\ \hline
        {\bf Sl. No.} & {\bf  Species Name} & {\bf Description} & {\bf Notation} \\ \hline
        1. & p53 & Unbound p53 protein & $x_1$ \\ \hline
        2. & MDM2 & Unbound MDM2 protein & $x_2$ \\ \hline
        3. & $MDM2\_p53$ & MDM2/p53 complex & $x_3$ \\ \hline
        4. & $MDM2\_mRNA$ & MDM2 messenger RNA & $x_4$ \\ \hline
        5. & $NO$ & Unbound Nitric Oxide  & $x_5$ \\ \hline
        6. & $NO\_MDM2$ & NO/MDM2 complex & $x_6$ \\ \hline        
\end{tabular}
\end{center}
\end{table}

Cellular processes are complex and stochastic or noise induced processes due to random molecular interaction in the system \cite{Rao} and system interaction with the environment \cite{Mca,Bla}. In stochastic model the molecular interaction is described by Master equation formalism to find the trajectories of molecular species variables in configuration space \cite{Mac,Gill,Gill1,Kam}. In this formalism, the time evolution of configurational probability $P({\bf x},t)$ of the state variable ${\bf x(t)}$ is constructed based on the assumption that firing of reaction/reactions within each time step is associated with decay and creation of molecular species \cite{Gill,Mac,Kam}. Normally, Master equation for complex system is very difficult to solve analytically, therefore one can use the stochastic simulation algorithm (SSA) (Gillespie algorithm) formulated from Master equation to compute the trajectory of each molecular species \cite{Gill,Gill1}. SSA is Monte carlo type numerical algorithm to simulate time evolution of chemical reactions system by taking every possible interaction in the system based on the question, which reaction fires at what time \cite{Gill}. 
\begin{figure}
\centering{\includegraphics[width=100mm]{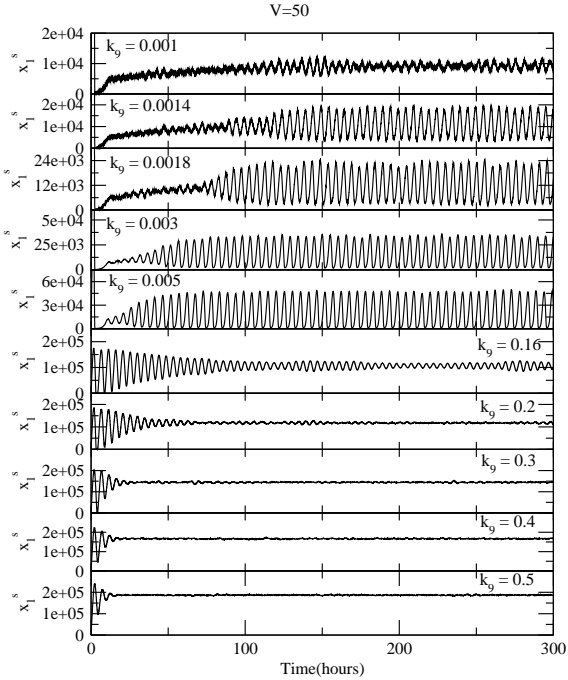}}
\caption{Temporal behaviors of p53 for different values of $k_9$ in Single cell by using Stochastic approach for V=50.}
\end{figure}
Further, one can simplify the Master equation based on two realistic assumptions: first on the small time interval limit of any two consecutive states along the system dynamics allowing propensity functions approximately unchange and second, on large molecular population limit with large time interval limit to propensity function of each reaction channels to be larger than one allowing to approximate the molecular distribution to normal distribution \cite{Gill1}. This let the Master equation to reduce to simpler Chemical Langevin equations (CLE). For our system, we have following CLEs,
\begin{eqnarray}
\label{cle}
\frac{d}{dt}{\bf x}(t)&=&{\bf M}(x_1,x_2,\dots,x_6;t)\nonumber\\
&&+\frac{1}{\sqrt{V}}{\bf G}(x_1,x_2,\dots,x_6,\xi_{1},\dots,\xi_{19};t)
\end{eqnarray}
where, ${\bf G}=[g_1,g_2,\dots,g_6]^T$ is a vector which is the contribution of noise: $g_1=-\sqrt{k_3x_1x_2}\xi_1+\sqrt{k_4x_3}\xi_2+\sqrt{k_8}\xi_3$, $g_2=\sqrt{k_2x_4}\xi_4-\sqrt{k_3x_1x_2}\xi_5+\sqrt{k_4x_3}\xi_6+\sqrt{k_5x_3}\xi_7-\sqrt{k_6x_2}\xi_8-\sqrt{k_{10}x_5x_2}\xi_9$, $g_3=\sqrt{k_3x_1x_2}\xi_{10}-\sqrt{k_4x_3}\xi_{11}-\sqrt{k_5x_3}\xi_{12}$, $g_4=\sqrt{k_1x_1}\xi_{13}-\sqrt{k_7x_4}\xi_{14}$, $g_5=\sqrt{k_9}\xi_{15}-\sqrt{k_{10}x_5x_2}\xi_{16}+\sqrt{k_{11}x_6}\xi_{17}-\sqrt{k_{12}x_5}\xi_{18}$, $g_6=\sqrt{k_{10}x_5x_2}\xi_{19}-\sqrt{k_{11}x_6}\xi_{20}$.
$V$ is the system size and $\xi_i$, $i=1,2,\dots,20$ are random noise parameters which are given by, $\langle \xi_i(t)\xi_j(t^\prime)\rangle=\delta_{ij}\delta(t-t^\prime)$.

\begin{table}
\begin{center}
{\bf Table 2 List of Chemical Reactions, Kinetic laws and their rate contants } 
 \begin{tabular}{|l|p{1.5cm}|p{1.5cm}|p{2.0cm}|p{1.2cm}|}
 \hline \multicolumn{5}{}{} \\ \hline

        {\bf Sl. No.} & {\bf Reaction } &  {\bf Kinetic laws} & {\bf Values of rate constant}&{\bf References}\\ \hline
        1  & $x_1\stackrel{k_1}{\longrightarrow}x_1+x_4$ &  $k_1 x_1 $ & $1.0\times 10^{-4} sec^{-1}$ & \cite{Pro,Fin}.\\ \hline
        2  & $x_4\stackrel{k_2}{\longrightarrow}x_4+x_2$ &  $k_2 x_4 $ & $4.95\times10^{-4} sec^{-1}$ & \cite{Pro,Fin}. \\ \hline
        3  & $x_1+x_2\stackrel{k_3}{\longrightarrow}x_3$ &  $k_3 x_1 x_2 $ & $11.55\times 10^{-4} mol^{-1} sec^{-1}$ & \cite{Pro}. \\ \hline
        4  & $x_3\stackrel{k_4}{\longrightarrow}x_1+x_2$ &  $k_4 x_3 $ &$11.55\times 10^{-6} sec^{-1}$ & \cite{Pro,Fin}. \\ \hline       
        5  & $x_3\stackrel{k_5}{\longrightarrow}x_2$ &  $ k_5 x_3 $ & $8.25\times 10^{-4} sec^{-1}$ & \cite{Pro}. \\ \hline
        6  & $x_2\stackrel{k_6}{\longrightarrow}\phi$ & $k_6 x_2 $ & $4.33\times 10^{-4} sec{-1}$ & \cite{Pro}. \\ \hline
        7  & $x_4\stackrel{k_7}{\longrightarrow}\phi$ & $k_7 x_4 $ & $1.0\times 10^{-4} sec^{-1}$ & \cite{Pro,Fin}.\\ \hline
        8  & $\phi\stackrel{k_8}{\longrightarrow}x_1$ &  $k_8$ & $0.78 sec^{-1}$ & \cite{Pro}.\\ \hline
        9  & $\phi\stackrel{k_{9}}{\longrightarrow}x_5$ & $k_{9}$ & $1\times 10^{-2} mol^{-1} sec^{-1}$ & \cite{Xin,Jah}. \\ \hline
        10 & $x_5+x_2\stackrel{k_{10}}{\longrightarrow}x_6$ & $k_{10} x_5 x_2$ & $1\times 10^{-3} mol^{-1} sec^{-1}$ & \cite{Jah}.\\ \hline
        11 & $x_6\stackrel{k_{11}}{\longrightarrow}x_5$ & $k_{11} x_6$ & $3.3\times 10^{-4} sec^{-1}$ & \cite{Xin,Jah}.\\ \hline
        12 & $x_5\stackrel{k_{12}}{\longrightarrow}\phi$ & $k_{12} x_5$ & $1 \times 10^{-3} sec^{-1}$ &  \cite{Xin,Jah}.\\ \hline      
\end{tabular}
\end{center}
\end{table}

\subsection{\bf Systems interacting in a three dimensional network}

Consider $N$ number of systems arranged in a regular three dimensional array (Fig. 2) such that the dynamics of the state vector of the system at $(i,j,k)$ position is given by,
\begin{eqnarray}
\label{net}
\frac{d}{dt}{\bf x^{[ijk]}}(t)={\bf F^{[ijk]}}(x_1^{[ijk]},x_2^{[ijk]},\dots,x_6^{[ijk]};\xi_1,\dots;t)
\end{eqnarray}
where, ${\bf F^{[ijk]}}={\bf M^{[ijk]}}+\frac{1}{\sqrt{V}}{\bf G^{[ijk]}};i,j,k=1,2,\dots,L$, with ${\bf M^{[ijk]}}$ and ${\bf G^{[ijk]}}$ have similar functional form as given in equations (\ref{det}) and (\ref{cle}). The equation (\ref{net}) recovers deterministic dynamics if $V\rightarrow\infty$ with $\{\xi\}\rightarrow 0$. 
\begin{figure}
\centering{\includegraphics[width=80mm]{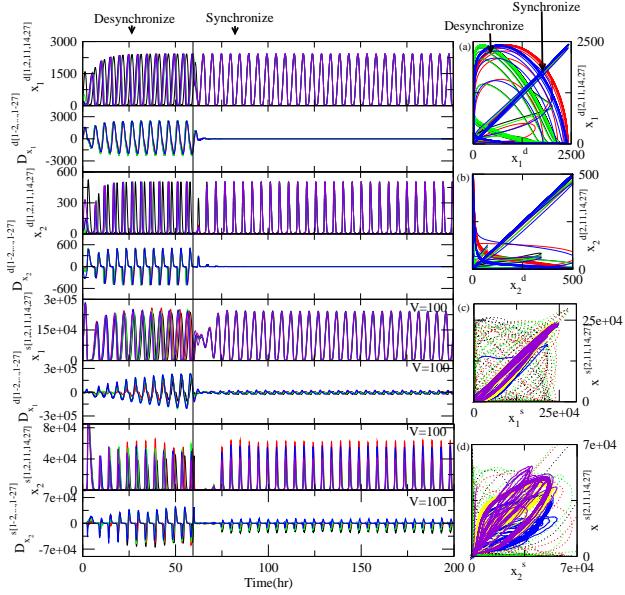}}
\caption{Synchronization of p53 and MDM2 dynamics interacting in three dimentional array shown.}
\end{figure}
Now let us allow the systems to interact with nearest neighbour i.e. for any system defined by ${\bf x^{[ijk]}}$ will have six nearest neighbours allowed to interact, namely, ${\bf x^{[i+1,jk]}}$, ${\bf x^{[i-1,jk]}}$, ${\bf x^{[ij+1,k]}}$, ${\bf x^{[ij-1,k]}}$, ${\bf x^{[ijk+1]}}$ and ${\bf x^{[ijk-1]}}$. If we consider bidirectional diffusive coupling among the systems via $x_5$ (NO), then in this interacting systems picture one can write any $(ijk)th$ system as,
\begin{figure}
\centering{\includegraphics[width=100mm]{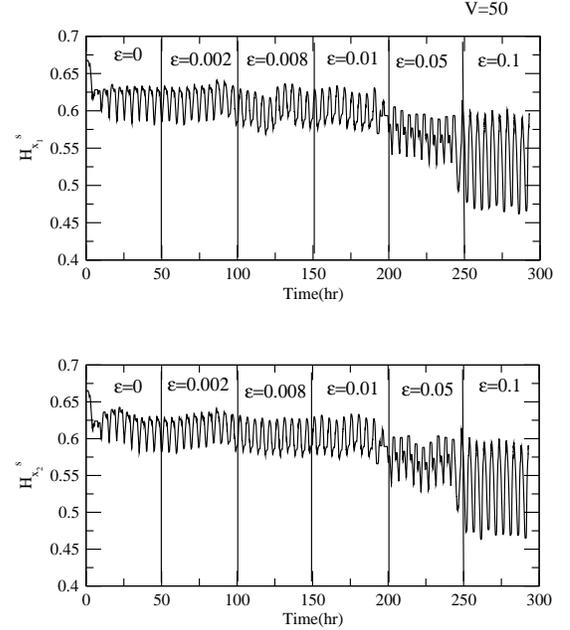}}
\caption{Permutation entropy spectrum for different values of $\epsilon$ for V=50.}
\end{figure}
\begin{eqnarray}
\label{ode}
\frac{d}{dt}{\bf x^{[ijk]}}(t)&=&{\bf F^{[ijk]}}\left(x_1^{[ijk]},x_2^{[ijk]},\dots,x_6^{[ijk]};\xi_1,\dots;t\right)\nonumber\\
&&+\sum_{i^\prime=-1}^{1}\epsilon^{[i,i+i^\prime]}\left( x_5^{[i+i^\prime,jk]}-x_5^{[ijk]} \right)\nonumber\\
&&+\sum_{j^\prime=-1}^{1}\epsilon^{[j,j+j^\prime]}\left( x_5^{[ij+j^\prime,k]}-x_5^{[ijk]} \right)\nonumber\\
&&+\sum_{k^\prime=-1}^{1}\epsilon^{[k,k+k^\prime]}\left( x_5^{[ijk+k^\prime]}-x_5^{[ijk]} \right)
\end{eqnarray}
where, $\{\epsilon\}$ are coupling constants. The extra terms in the equation (\ref{ode}) due to coupling are non-zero only for ODE of $x_{5}^{[ijk]}$. We also consider periodic boundary conditions i.e. ${\bf x^{[i+L,jk]}}={\bf x^{[ijk]}}$, ${\bf x^{[ij+L,k]}}={\bf x^{[ijk]}}$ and ${\bf x^{[ijk+L]}}={\bf x^{[ijk]}}$. 

\subsection{\bf Permutation entropy: measurement of complexity}
\begin{figure}
\centering{\includegraphics[width=100mm]{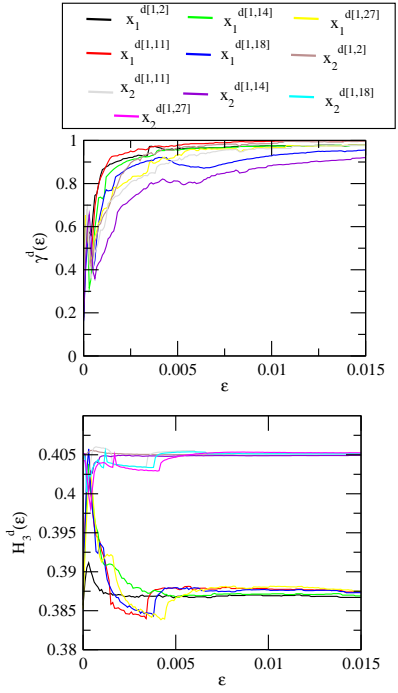}}
\caption{The plots of order parameter $\gamma^d$ and permutation entropy $H_3^d$ as a function of $\epsilon$ for deterministic case.}
\end{figure}
Consider a time series data (either stochastic or deterministic) $x(t)$ which can be mapped onto a long symbolic sequence of length $N$ given by, $x(t)=\{x_1,x_2,\dots,x_N\}$, then permutation entropy, $H$ of this data can be calculated as follows \cite{Ban,Cao}. Let us take a window sequence $W$ of size $L$ which is allowed to slide along the data sequence $x(t)$ with maximum overlap. This window sliding procedure allows $x$ to make $M$ equal partitions: $x=\{W_1,W_2,\dots,W_M\}$. Now to calculate any jth permutation entropy $H_j$ of window sequence $W_j=\{x_{j+1},x_{j+2},\dots,x_{j+L}\}$ a short sequence of embedded dimension $r$: $S_i=\{x_{i+1},x_{i+2},\dots,x_{i+r}\}$ is defined with all possible inequalities of dimension $r$, and mapping the inequalities along $W_j$ in accending order to obtain probabilities of occurrences of each inequalities ($p_j:j=1,2,...$). Since only $Q$ out of $r!$ permutations are distinct, the normalized permutation entropy, $H_j(r)=-\frac{1}{ln(r!)}\sum_{i=1}^{Q}p_ilnp_i$ can be calculated, where, $0\le H_j(r)\le 1$. Similarly permutation entropies of each window sequences in $x$ can be calculated, and thus permutation entropy spectrum of the time series data $x(t)$ is obtained as, $H(t)=\{H_i:i=1,2,\dots,M\}$. In stochastic systems generally noise enhances $H(t)$ that leads to increase in complexity in the dynamics, however there are cases where noise reduces $H(t)$ value \cite{Ban}. But if the strength of the noise is small enough, it does not cause significant change in complexity in the dynamics \cite{Ban}. This gives us a notion that noise has an important impact on $H(t)$ in stochastic dynamics.
\begin{figure}
\centering{\includegraphics[width=100mm]{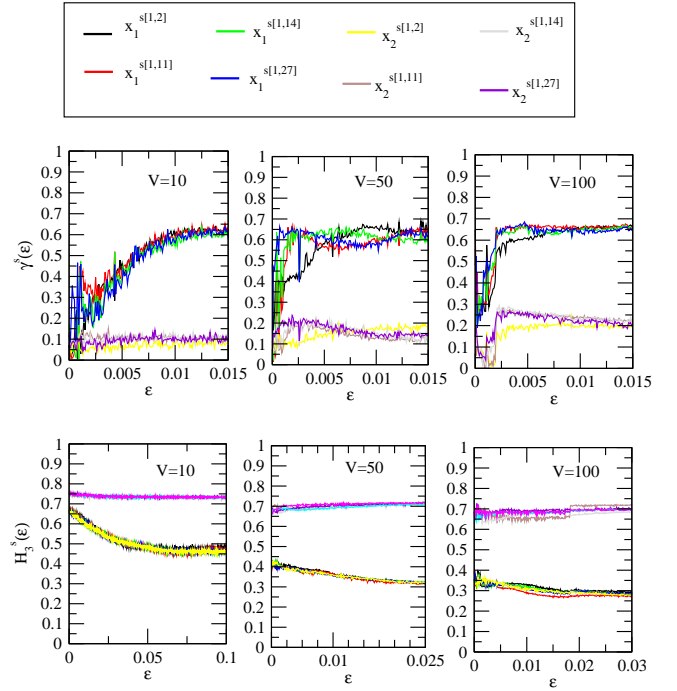}}
\caption{The plots of order parameter $\gamma^s$ and permutation entropy $H_3^s$ as a function of $\epsilon$ for different values of V=10,50,100 for stochastic case.}
\end{figure}
\subsection{\bf Detection and measurement of synchronization rate}

The concept of permutation entropy technique can be used to detect synchrony and synchronization rate of two coupled systems \cite{Liu}. The technique is used by defining the coupled system as $X(t)=[x^{[1]},x^{[2]}]^T$, where, the superscripts 1 and 2 indicate the two coupled subsystems coupled via a coupling parameter $\epsilon$. The permutation entropy spectra of the two subsystems are given by, $H^{[1]}(t)=\{H_1^{[1]},H_2^{[1]},\dots,H_M^{[1]}\}$ and $H^{[2]}(t)= \{H_1^{[2]},H_2^{[2]},\dots,H_M^{[2]}\}$ respectively. Then a correlation like function $C(x^{[j]}):j=1,2$ can be defined as,
\begin{eqnarray}
C(x^{[j]})&=&1~~~~H_{i}^{[j]}\rangle H_{i-1}^{[j]},~~~\forall i:i=1,2,\dots,M\nonumber\\
&=&-1~~~~otherwise
\end{eqnarray}
Now the order parameter $\gamma$ for the coupled system is calculated as follows,
\begin{eqnarray}
\label{gama}
\gamma=\langle C(x^{[1]})C(x^{[2]})\rangle
\end{eqnarray}
where, $\langle\dots\rangle$ is time average. If one calculate $\gamma(\epsilon)$ as a function of coupling parameter $\epsilon$, then one can detect that the two subsystems are uncoupled if $\gamma=0$, but they are synchronized if $\gamma=1$ \cite{Liu}. The rate of synchronization can be measured from the values of $\gamma$ of the coupled system such that, for weak synchronization $\gamma\rightarrow 0$, and for strong synchronization $\gamma\rightarrow 1$.
\begin{figure}
\centering{\includegraphics[width=100mm]{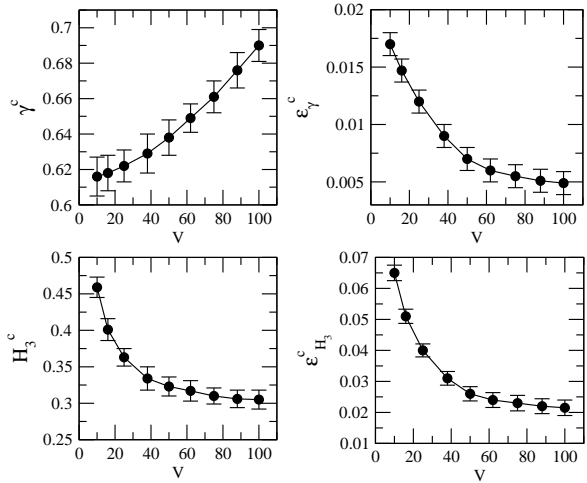}}
\caption{The variation of $\gamma^c$, $\epsilon_\gamma^c$, $H_3^c$ and $\epsilon_{H_3}^c$ as a function of system size.}
\end{figure}

Another way of measuring synchronization rate of the coupled system qualitatively is to measure a distance function parameter, $D_k^{[1,2]}(t)=|x_k^{[1]}(t)-x_k^{[2]}(t)|~:\forall k;k=1,2,\dots,N$ \cite{Pec,Ram,Ros,Ros1}. The two subsystems are in (i) synchronous state if $D_k^{[1,2]}(t)\rightarrow 0$, (ii) desynchronized state if $D_k^{[1,2]}(t)$ fluctuates randomly, and (iii) transition state if the rate of fluctuation is about a constant value, that is, $0~\langle~ D_k^{[1,2]}(t)~\langle~ fluctuation~ (in~desynchronized~ case)$.

The rate of synchronization can also be detected qualitatively by two dimensional recurrence plot of the corresponding variables of the coupled systems \cite{Pec}. The two subsystems are uncoupled if the points in plot are distributed randomly away from the diagonal. However, if the two subsystems start coupling each other then the points in the plot start concentrating along the diagonal. The rate of synchronization is indicated by the rate of broadening of the points along the diagonal. If the two systems are strongly synchronized the points are just aligned along the diagonal, however, if the two systems are weakly synchronized, the points are scattered away from the diagonal showing a broad diagonal line.

\subsection{\bf Simulation details}

Single system dynamics is investigated at various stress conditions induced by $NO$ deterministically and stochastically to understand how does the system undergo different states under these stress conditions. We used 4th order Runge Kutta method of numerical integration to solve deterministic as well as stochastic coupled differential equations \cite{Pre} and for stochastic system, $Gillespie's$ Stochastic Simulation Algorithm (SSA) \cite{Gill,Gill1} is used to trace the trajectory of the system dynamics. We wrote our own code in java for simulation purpose \cite{ Her}.

We then consider a network of identical stress systems (27 systems) arranged in a three dimensional array as shown in Fig. 2. Each system is maintained at a certain stress condition induced by diffusing $NO$ by keeping the system at a particular value of $k_{NO}$. We allow bidirectional diffusive coupling of each system with nearest neighbour systems only via NO, and we take $\epsilon^{ijk}=\epsilon;\forall i,j,k=1,2,\dots,L$. Periodic boundary conditions are applied in simulating reaction sets of the systems in the network.

\section{Results and Discussion}

We present our results of $p53-MDM2$ dynamics induced by $NO$ and interaction among such identical systems in a certain topology.

\subsection{\bf Driving $p53$ at various states by $NO$}

We present deterministic results of $p53$ dynamics at various stress conditions indicated by different $k_{NO}~(k_9)$ (different values of $NO$ production rate in the system) values [$0.0012-0.5$] as shown in Fig. 3 by numerically solving equation (\ref{det}). All the parameter values for the simulation are shown in the Table 2. At low value of $k_{NO}$ (=0.0012) the $p53$ level maintains at low value (nearly normal condition). If $T_t$ is critical transition time which can be defined as the time which separates oscillatory regime and oscillation death (stabilized or oscillation death) regime, we found that $T_t\rightarrow 0$ as $k_{NO}$ goes from small to moderate value (0.0012 to 0.0065) indicating switching $p53$ state from normal to activated state induced by $NO$. For $k_{NO}$ values in the range [0.005-0.155], $T_t\rightarrow\infty$ indicating sustained oscillation regime where activation of $p53$ is maximized. If we further increase $k_{NO}$ value ([0.162-0.5]), $T_t\rightarrow 0$ showing the system move to stabilized regime indicating excess $NO$ induce the system to apoptotic state \cite{Bar,San,Jah}. 

The stochastic results are obtained by simulating the reaction set listed in Table 2 using SSA \cite{Gill} with the same parameter values used in deterministic case for $V=10,50,100$ and similar behaviour as in deterministic case is obtained with fluctuation in the dynamics due to intrinsic noise (Fig. 4). The stochastic results show that $p53$ responds somewhat faster to the stress exerted by $NO$ than that in deterministic case probably due to intrinsic noise in the system. 

\subsection{\bf Synchronization induced by $NO$}
We now present the results of synchronization of identical stress systems via $NO$. Each system in the network is kept under certain stress ($k_{NO}=0.04$). We take coupling constant to be the same ($\epsilon=0.01$) and coupling is switched on at time 60 hours (Fig. 5). The deterministic and stochastic behaviours of $p53$ ($x_1^{[ijk]}$) and $MDM2$ ($x_2^{[ijk]}$) for system at (111) coupled with nearest neighbour systems at (011), (211), (101), (121), (110) and (112) respectively are shown in Fig. 5 (the first four upper panels are for deterministic and the last four lower panels are for stochastic case for $V=100$). The plots clearly show three distinct regimes, namely, desynchronized, transition and synchronized regimes respectively. The claim is then supported by the behaviour of $D^{[ijk,i^\prime j^\prime k^\prime]}(t):\forall i^\prime=i\pm1,j^\prime=j\pm1,k^\prime=k\pm1$ identifying desynchronized ($D^{[ijk,i^\prime j^\prime k^\prime]}$ fluctuates randomly as a function of time), transition ($D^{[ijk,i^\prime j^\prime k^\prime]}$ fluctuates a little about a constant) and synchronized (fluctuation of $D^{[ijk,i^\prime j^\prime k^\prime]}$ about a constant is minimum) regimes both in deterministic and stochastic systems. However, the fluctuation in $D^{[ijk,i^\prime j^\prime k^\prime]}$ in synchronized regime in deterministic system is much smaller (almost zero) than that of in stochastic system. This exhibits the significant hindarence behaviour of noise in stochastic system. Further this claim is supported by the two dimensional reccurrence plot of each pair of coupled systems as shown in right panels of Fig. 5. The three regimes can be identified clearly from the plots where, for desynchronized regime the points in the plots are randomly distributing away from the diagonal, for transition regime the rate of distribution of points away from the diagonal is comparatively smaller than in desynchronized case and for synchronized regime the points are concentrated along the diagonal.

We calculate permutation entropy spectra of the time series data of $x_1^{[ijk]}$ and $x_2^{[ijk]}$ in the network for different coupling constants (Fig. 6). In our calculation we take the value of embedded dimension $r=3$ and size of the window $L=1000$. The dynamics of $H$ of $p53$ and $MDM2$ for different values of $\epsilon$ are shown in Fig. 6 and is found that $H$ for fixed value of $\epsilon$ exhibits oscillatory behaviour about a constant mean value. However, as $\epsilon$ increases $H$ start decreasing. This could be due to increase in $\epsilon$ leads the coupled systems to more ordered direction because of the information processing via MDM2 and then to p53.

Since the order parameter $\gamma$ can be used as an indicator of the degree of synchronization \cite{Liu}, we calculated this parameter to characterize the synchronization among the coupled systems in the network. The deterministic results of $\gamma^d$ and permutation entropy spectrum for p53 ($x_1^{[ijk]}$) and MDM2 ($x_2^{[ijk]}$) as a function of $\epsilon$ are shown in Fig. 7. The $\gamma^d$ dependence on $\epsilon$ (Fig. 7 upper panel) shows that all far and near systems achieve synchronization approximately after $\epsilon_c=0.0027\pm 0.0021$, and below this value the systems show weak synchronization (uncouple when the value of $\gamma^d$ is significantly low). It also shows that $\gamma^d$ for MDM2 reach synchronization at larger value of $\epsilon$ and at smaller value of $\gamma$ as compared to p53. It is also seen in the permutation entropy spectrum that $H_3^d$ for p53 decreases as $\epsilon$ increases and remain almost constant after $\epsilon_c$, however,$H_3^d$  for MDM2  increases as $\epsilon$ increases and remains constant after the same value $\epsilon_c$. 

\subsection{\bf The role of NO}
$NO$ in the system plays two contrast roles, first, it induces stress to the system causing increase in disorderness, and second, it behaves as synchronizing agent inducing more orderness to the system. The direct interaction of diffusing $NO$ with $MDM2$ leads to decrease in $MDM2$ concentration level but increase in $p53$ level (Fig. 7). For small values of $\epsilon$, the role of stress induced in the system could be more significant than the role of inducing synchronization to the system. The direct interaction of $NO$ with $MDM2$ may originate more disorder in $MDM2$ resulting $H_3^d$ of $MDM2$ to increase (Fig. 7). After significant amount of $NO$ is diffused (after $\epsilon_c$) optimal amount of orderness may be maintained such that $H_3^d$ of $MDM2$ becomes constant as a function of $\epsilon$. However, decrease in $p53$ due to indirect interaction with $NO$ via $MDM2$ may establish decrease in $H_3^d$ of $p53$ as a function of $\epsilon$. But since $p53$ level will also attain optimal orderness due to $MDM2$, $H_3^d$ of $p53$ also becomes constant after $\epsilon_c$ (Fig. 7).

\subsection{\bf Effect of noise}
The stochastic results of $\gamma^s$ and $H_3^s$ for p53 and MDM2 as a function of $\epsilon$ for three different values of system sizes $V=$10, 50 and 100 are presented in Fig. 8 for the same parameter values used in deterministic case. We found similar behaviours of $\gamma^s$ and $H_3^s$ as in the case of deterministic ones. However, synchronization among the systems is achieved at much lower values of $\gamma^s$ ($\sim 0.68\pm 0.07$) and the value of $\epsilon_c$ is dependent on system size (for V=10, $\epsilon_c=0.01\pm0.003$; V=50, $\epsilon_c=0.005\pm0.001$; V=100, $\epsilon_c=0.003\pm0.0005$). The $\gamma^s$ values of p53 and MDM2 are significantly different as shown in the plots. If we define $\gamma^c$ as the critical value of $\gamma^s$ below which the systems are in weakly synchronized regime and above which the systems exhibit strongly synchronized regime, difference between $\gamma^c$ for p53 and MDM2 is quite large ($\sim 0.52$) and remains constant approximately as $V$ changes. The separation between $H_3^s$ of p53 and MDM2 increases as $V$ increases. It may be because of the fact that as $V$ increases the strength of molecular noise in the system will decrease which leads to the system to orderness direction.

We then study the behaviour of $\gamma^c$ as a function of $V$ (Fig. 9 upper panel first) and found that $\gamma^c$ increases as $V$ increases and becomes constant for large values of $V$ (deterministic limit). $\gamma^c$ varies with $V$ by the functional form, $\gamma^c=\alpha V^\beta +\delta$ separating weak and strong synchronization regimes. Also $\epsilon_\gamma^c$ found to decay exponentially as a function of $V$ i.e. $\epsilon_\gamma^c=ae^{-bV}$ (Fig. 9 upper panel second). $H_3^c$ of p53 and MDM2 also found to decrease exponentially as a function of $V$, $H_3^c=ue^{-wV}$ (Fig. 9 lower panel first). Further $\epsilon_{H_3}^c$ decreases as $V$ increases with functional form, $\epsilon_{H_3}^c=ce^{-dV}$ (Fig. 9 lower panel second).

\subsection{\bf Stability analysis}

The deterministic stationary state of the system can be obtained by putting $\frac{d}{dt}{\bf x^{[ijk]}}=0$ in equation (\ref{net}) giving six stationary equations from which stationary values of the variable ${\bf x^{[ijk]*}}=(x_1^{[ijk]*},x_2^{[ijk]*},\dots,x_6^{[ijk]*})^T$ can be obtained. The stationary equation for $x_5^{[ijk]*}$ is given by,
\begin{eqnarray}
&&M_n^{[ijk]}\left(x_1^{[ijk]*},x_2^{[ijk]*},\dots,x_6^{[ijk]*}\right)=0;\forall n,n\ne 5\\
&&x_5^{[ijk]*}\left(k_9-k_{11}-\epsilon^{[i,i-1]}-\epsilon^{[i,i+1]}\right)\nonumber\\
&&-x_5^{[ijk]*}\left(\epsilon^{[j,j-1]}-\epsilon^{[j,j+1]}-\epsilon^{[k,k-1]}-\epsilon^{[k,k+1]}\right)\nonumber\\
&&+\epsilon^{[i,i-1]}x_5^{[i-1,jk]*}+\epsilon^{[i,i+1]}x_5^{[i+1,jk]*}+\epsilon^{[j,j-1]}x_5^{[i,j-1,k]*}\nonumber\\
&&+\epsilon^{[j,j+1]}x_5^{[i,j+1,k]*}+\epsilon^{[k,k-1]}x_5^{[ij,k-1]*}\nonumber\\
&&+\epsilon^{[k,k+1]}x_5^{[ij,k+1]*}=0
\end{eqnarray}
Simplifying the coupling constants by taking, $\epsilon^{[i,i-1]*}=\epsilon^{[i,i+1]*}=\epsilon^{[j,j-1]*}=\epsilon^{[j,j+1]*}=\epsilon^{[k,k-1]*}=\epsilon^{[k,k+1]*}=\epsilon$, we can rewrite the above stationary equations as,
\begin{eqnarray}
\label{s1}
&&M_n^{[ijk]}\left(x_1^{[ijk]*},x_2^{[ijk]*},\dots,x_6^{[ijk]*}\right)=0;\forall n,n\ne 5\\
\label{s2}
&&x_5^{[ijk]*}=\frac{\epsilon}{6\epsilon+k_{11}-k_9}\left(\sum_{\langle ii^\prime\rangle,\langle jj^\prime\rangle,\langle kk^\prime\rangle}x_5^{[i^\prime j^\prime k^\prime]*}\right)
\end{eqnarray}
where, $\langle\dots\rangle$ denotes the nearest neightbour interaction. If we consider coupling of two consecutive systems $[ijk]$ and $[i-1,jk]$ in the network then we have the following stationary equations of these two coupled systems as,
\begin{eqnarray}
\label{sc1}
&&M_n^{[ijk]}\left(x_1^{[ijk]*},x_2^{[ijk]*},\dots,x_6^{[ijk]*}\right)=0\\
\label{sc2}
&&M_n^{[i-1,jk]}\left(x_1^{[i-1,jk]*},x_2^{[i-1,jk]*},\dots,x_6^{[i-1,jk]*}\right)=0\\
\label{sc3}
&&x_5^{[ijk]*}=\frac{\epsilon}{6\epsilon+k_{11}-k_9}\left(\sum_{\langle ii-1\rangle,\langle jj^\prime\rangle,\langle kk^\prime\rangle}x_5^{[i j^\prime k^\prime]*}\right)\\
\label{sc4}
&&x_5^{[i-1,jk]*}=\frac{\epsilon}{6\epsilon+k_{11}-k_9}\left(\sum_{\langle i-1,i\rangle,\langle jj^\prime\rangle,\langle kk^\prime\rangle}x_5^{[i-1, j^\prime k^\prime]*}\right)\nonumber\\
\end{eqnarray}
where, $\forall n,n\ne 5$. Subtracting the equations (\ref{sc3}) and (\ref{sc4}) we get the following equation,
\begin{eqnarray}
\Delta x_5^{[i,i-1,jk]*}=\frac{\epsilon}{7\epsilon-k_9-k_{11}}\Gamma
\end{eqnarray}
where, $\Delta x_5^{[i,i-1,jk]*}=x_5^{[ijk]}-x_5^{[i-1,jk]}$ which is the net NO diffused between the two systems in the network. The function $\Gamma$ is given by,
\begin{eqnarray}
\Gamma=\sum_{\langle i+1,i-2\rangle jk}\left(x_5^{[ijk]*}-x_5^{[i-1,jk]*}\right)
\end{eqnarray}
Solving equations (\ref{sc1}) and (\ref{sc2}) we have the following equation for p53,
\begin{eqnarray}
\frac{\Delta x_1^{[i,i-1,jk]*}}{x_1^{[i-1,jk]*}}\sim\frac{1}{2k_9}\left(\frac{\epsilon}{k_9-k_{11}-7\epsilon}\right)\left(\frac{\Gamma}{x_1^{[ijk]*}}\right)
\end{eqnarray}
where, $\frac{\Delta x_1^{[i,i-1,jk]*}}{x_1^{[i-1,jk]*}}=\frac{x_1^{[ijk]*}-x_1^{[i-1jk]*}}{x_1^{[i-1,jk]*}}$ which is average change (fractional change) of $x_1^{[ijk]*}$. Since $\frac{\Delta x_1^{[i,i-1,jk]*}}{x_1^{[i-1,jk]*}}=\frac{x_1^{[ijk]*}-x_1^{[i-1jk]*}}{x_1^{[i-1,jk]*}}\propto\frac{\Gamma}{x_5^{[ijk]*}}$, the average rate of change $x_1^{[ijk]*}$ to reach the stable state is proportional to the average change of $x_5^{[ijk]*}$.

Similarly solving equations (\ref{sc1}) and (\ref{sc2}) we get the expression for average rate of change of $x_2^{[ijk]}$ as follows,
\begin{eqnarray}
\frac{\Delta x_1^{[i,i-1,jk]*}}{x_1^{[i-1,jk]*}}&\sim&\frac{\frac{1}{2k_9}\left(\frac{\epsilon}{k_9-k_{11}-7\epsilon}\right)\left(\frac{\Gamma}{x_1^{[ijk]*}}\right)}{1-\frac{1}{2k_9}\left(\frac{\epsilon}{k_9-k_{11}-7\epsilon}\right)\left(\frac{\Gamma}{x_1^{[ijk]*}}\right)}\nonumber\\
&\sim&\frac{\frac{\Delta x_1^{[i,i-1,jk]*}}{x_1^{[i-1,jk]*}}}{1-\frac{\Delta x_1^{[i,i-1,jk]*}}{x_1^{[i-1,jk]*}}}
\end{eqnarray}
The fractional change of $x_2^{[ijk]*}$ is also found to be proportional to $\frac{\Delta x_1^{[i,i-1,jk]*}}{x_1^{[i-1,jk]*}}$.

\section{Conclusion}

Cellular function at molecular level under stress condition induced by NO exhibits different behaviour as compared to normal cell function and allows the cell to go to different states. The single system of p53-MDM2-NO network we studied shows three different distinct states, namely, fixed point oscillation at low values of NO concentration level (normal state), damped oscillation for moderate values of NO level (mixed state of small activated and stabilized state), sustain oscillation for large values of NO concentration level (activated state) and fixed point oscillation for excess NO level in the system (probably apoptosis). This switching of p53 behaviour at different states vis stress inducing molecules is believed to be reversible if the amount of stress induced is not excess. 

The systems interacting in a certain network topology will give us some important phenomena how systems in the network process signal. The network of systems arranged in a three dimensional regular array are allowed to interact with certain coupling constant with nearest neighbour interaction and the results show that the way the systems interact from nearest to far systems do not show significant difference. This means that once the coupling is switched on the systems process signal almost instaneously. The permutation entropy calculations of this model also show that the rate correlation among the system increases as the value of coupling constant is increased. We also found that the correlation of the systems becomes almost constant after one increase the value of coupling constant larger that certain value. The information contained in p53 and MDM2 dynamics is found to be different even if the proteins are in the same system. The reason is that the coupling molecule NO interacts with MDM2 first and passes the impact to p53 indirectly via MDM2. The deterministic and stochastic results show similar behaviours except the impact of noise in the system dynamics. Our results show that noise acts as a disturbing behaviour on synchronization. The value of coupling constant needed to synchronize the coupled systems significantly decreases as system size increases and becomes constant which could the approximate thermodynamic limit. However, one has to take huge number of systems to see robustness of this interaction and system level understanding.







\begin{thebibliography}{}
\bibitem{Vou} Vousden KH, Lu X (2002) live or let die: the $cell's$ response to p53. Nature 2:594-604.
\bibitem{Lev}Levine AJ (1997) p53, the Cellular Gatekeeper for Growth and Division. Cell 88:323-331.
\bibitem{Vog} Vogelstein B, Lane D, Levine AJ (2000) Surfing the p53 network. Nature 408:307-310. 
\bibitem{Lane} Lane DP (1992) Cancer. p53, guardian of the genome. Nature 358:15–16.
\bibitem{livi} LR Livingstone, A White, J Sprouse, E Livanos, T Jacks and TD Tlsty (1992) Altered cell cycle arrest and gene amplification potential accompany loss of wild-type p53. Cell 70, 923.
\bibitem{murr} F Murray-Zmijewski, D P Lane and J-C Bourdon (2006) p53/p63/p73 isoforms: an orchestra of isoforms to harmonise cell differentiation and response to stress. Cell Death and Differentiation 13, 962–972. 
\bibitem{serr} M Serrano, AW Lin, ME McCurrach, D Beach and SW Lowe (1997) Oncogenic ras Provokes Premature Cell Senescence Associated with Accumulation of p53 and p16INK4a. Cell 88, 593.
\bibitem{saka} K Sakaguchi, JE Herrera, S Saito, T Miki, M Bustin, A Vassilev, CW Anderson and E Appella (1998) DNA damage activates p53 through a phosphorylation–acetylation cascade. Genes and Dev. 12, 2831-2841.
\bibitem{yoni} E Yonish-Rouach, D Resnftzky, J Lotem, L Sachs, A Kimchi and M Oren (1991) Wild-type p53 induces apoptosis of myeloid leukaemic cells that is inhibited by interleukin-6. Nature 352, 345-347.
\bibitem{Bat}Bates S, Vousden KH (1996) p53 in signalling checkpoint arrest orapoptosis. Curr. Opin. Genet. Dev. 6:1–7.
\bibitem{Kub}Kubbutat MHG, Vousden KH (1998) Keeping an old friend undercontrol: regulation of p53 stability. Mol. Med. Today 4:250–256.
\bibitem{Cho}Choi J, Donehower LA (1999) p53 in embryonic development: maintaining a fine balance. Cell. Mol. Life Sci. 55:38–47.
\bibitem{Hau}Haupt Y, Maya R, Kazaz A, Oren M (1997) Mdm2 promotes the rapid degradation of p53. Nature 387:296-299.
\bibitem{Hon}Honda R, Tanaka H, Yasuda H (1997) Oncoprotein MDM2 is aubiquitin ligase E3 for tumor suppressor p53. FEBS Lett. 420:25–27.
\bibitem{Tao}Tao W, Levine AJ (1999) Nucleocytoplasmic shuttling of oncoproteinHdm2 is required for Hdm2-mediated degradation of p53. Proc. Natl. Acad.Sci. USA 96:3077–3080.
\bibitem{Wu}Wu XW, Bayle JH, Olson D, Levine AJ (1993) The p53-Mdm2 autoregulatory feedback loop. Genes Dev. 7:1126–1132.
\bibitem{She} Jin S, Levine AJ (2001) The p53 functional circuit, Journal of Cell Science 114:4139-4140.
\bibitem{Mee} Meek DW (1998) Multisite phosphorylation and the integration of stress signals at p53. Cell. Signal. 10:159–166.
\bibitem{Gu}Gu W, Roeder RG (1997) Activation of p53 sequence-specific DNA binding by acetylation of the p53 C-terminal domain. Cell 90:595–606.
\bibitem{Shaw}Shaw P, Freeman J, Bovey R, Iggo R (1996) Regulationof specific DNA binding by p53: evidence for a role of O-glycosylation and charged residues at the carboxy terminus. Oncogene 12:921–930.
\bibitem{Stern}Stern JE (2004) Nitric oxide and homeostatic control: an intercellular signalling molecule contributing to autonomic and neuroendocrine integration?. Progress in Biophysics and Mol. Bio. 84:197-215.
\bibitem{Zwe}Zweier  JL, Samouilov A, Kuppusamy (1999) Non-enzymatic nitric oxide synthesis in biological systems. Biochim.Biophys.Acta 1411:250–262.
\bibitem{Xin} Wang X, Michael D, Murcia GD, Oren M (2002) p53 Activation by Nitric Oxide Involves Down-regulation of Mdm2. journal of biological chemistry 277:15697–15702.
\bibitem{torr} J Torreilles (2001)  Nitric oxide: one of the more conserved and widespread signaling molecules.  Frontiers in Bioscience : a Journal and Virtual Library [2001, 6:D1161-72].
\bibitem{Gal} Lahav G, Rosenfeld N, Sigal A, Zatorsky NG, Levine AJ,  Elowitz MB, Alon U (2004) Dynamics of the p53-Mdm2 feedback loop in individual cells. Nat. Genet 36:147-150.
\bibitem{Bar} Bar-Or RL, Maya R, Segel LA, Alon U (2000) Generation of Oscillation by the p53-Mdm2 feedback loop: a theoretical and experimental study. Proc. Natl. Acad. Sci. USA 97:11250–11255.
\bibitem{San} Harris SL, Levine AJ (2005) The p53 pathway: positive and negative feedback loops. Oncogene 24:2899–2908.
\bibitem{Jah} Alam MJ, Devi GR, Ravins, Ishrat R, Agarwal SM, Singh RKB (2013) Switching p53 states by calcium: dynamics and interaction of stress systems. Mol. BioSyst. 9:508.
\bibitem{Nic} Schneiderhan N, Budde A, Zhang Y, Brune B (2003) Nitric oxide induces phosphorylation of p53 and impairs nuclear export. Oncogene 22:2857–2868.
\bibitem{Wan}  Wang X,  Zalcenstein A, Oren M (2003) Nitric oxide promotes p53 nuclear retention and sensitizes neuroblastoma cells to apoptosis byionizing radiation. Cell Death and Differentiation 10:468–476.
\bibitem{Wang1} Wang X, Michael D, de Murcia G, Oren M (2002) p53 Activation by nitric oxide involves down-regulation of Mdm2. J Bio. Chem. 277:15697-15702.
\bibitem{Sung} Lee SJ, Kim DC, Choi BH, Ha H, Kim KT (2006) Regulation of p53 by Activated Protein Kinase C- during Nitric Oxide-induced Dopaminergic Cell Death. J. Biol. chem. 281, 2215–2224.
\bibitem{yan} Yan S, Chen J, Wang R (2008) Towards anomalous diffusion with nonlinear interactions for Hamiltonian chaotic systems. Physica A 387, 1786–1798.
\bibitem{bres} Bressloff PC and Newby JM (2013) Stochastic models of intracellular transport. Rev. Mod. Phys. 85, 135-196.
\bibitem{yan1} Yan S, Sakata F, and Zhuo Y (2002) Features of statistical dynamics in a finite system. Phys. Rev. E 65 031111.
\bibitem{bo} Bo L and Shi-Wei Y (2011) Cellular Response to Irradiation. Commun. Theor. Phys. 55, 921–924.
\bibitem{bo1} Bo L and Yan S (2011) Nonlinear features in protein circuitry. Commun. Nonlinear Sci. Numer. Simulat. 16, 2957–2961.
\bibitem{bo2} Bo L, Shi-Wei Y and Yi-Zhao G (2011) Radiation-induced robust oscillation and non-Gaussian fluctuation. Chin. Phys. B 20, 128702.
\bibitem{bo3} Bo L, Yan S, Wang Q and Liu S (2011) Oscillatory expression and variability in p53 regulatory network. Physica D 240, 259–264.
\bibitem{zak} Zaks MA, Neiman AB, Feistel S, and Schimansky-Geier L (2003) Noise-controlled oscillations and their bifurcations in coupled phase oscillators. Phys. Rev. E 68, 066206.
\bibitem{zho} Zhou T, Chen L and Aihara K (2005) Molecular Communication through Stochastic Synchronization Induced by Extracellular Fluctuations. Phys. Rev. Lett. 95, 178103.
\bibitem{lind} Lindner B, GarcÃa-Ojalvob J, Neimand A, Schimansky-Geier L (2004) Effects of noise in excitable systems. Phys. Rep. 392, 321 – 424.
\bibitem{bo4} Bo L, Yan S and Wang Q (2011) Intrinsic noise and Hill dynamics in the p53 system. J. Theor. Biol. 269, 104–108.
\bibitem{yan2} Yan S, Wang Q and Liu S (2009) Towards a dynamical temperature of finite Hamiltonian systems. Physica A 388, 4943–4949.
\bibitem{yan4} Yan S and Zhuo Y (2006) A unified model for studying DNA damage-induced p53–Mdm2 interaction. Physica D 220, 157–162.
\bibitem{yan5} Yan S (2007) Negative feedback dynamics and oscillatory activities in regulatory biological networks. J. Biol. Syst., 15, 123
\bibitem{sheng}Sheng-Jun L, Qi W, Bo L, Shi-Wei Y (2011) Fumihiko Sakata, Noise transmission and delay-induced stochastic oscillations in biochemical network motifs. Chin. Phys. B., 20 (12): 128703.
\bibitem{wangq}Wang Q, Liu B and YAN S (2011) Oscillatory dynamics induced by multi-delays in gene expression. Advs. Complex Syst. 14, 451.
\bibitem{bo5} Bo L, Yan S and Gao X (2011) Noise Amplification in Human Tumor Suppression following Gamma Irradiation. PLoS ONE 6, e22487.
\bibitem{Pro} Proctor CJ, Gray DA (2008) Explaining oscillations and variability in the p53-Mdm2 system. BMC Systems Biol.  2:75.
\bibitem{Fin} Finlay CA (1993) The Mdm2 oncogene can overcome wild-type p53 suppression of transformed cell growth. Mol. Cell. Bio. 13:301.
\bibitem{Mol} Moll UM, Petrenko O (2003) The MDM2-p53 interaction.  Mol. Cancer Research  1:1001.
\bibitem{Moma}Momand J, Wu HH,  Dasgupta G (2000) MDM2--master regulator of the p53 tumor suppressor protein. Gene 242:15-29.
\bibitem{Wood}Wood J, Garthwaite J (1994) Models of the diffusional spread of nitric oxide: implications for neural nitric oxide signalling and its pharmacological properties. Neuropharmacology 33:1235-1244.
\bibitem{Sci}Schonhoff CM,Daou MC,Jones SN,Schiffer CA, Ross AH (2002) Nitric oxide-mediated inhibition of Hdm2-p53 binding. Biochemistry 41:13570-13574.
\bibitem{Chen} Chen J,Lin J, Levine AJ (1995) Regulation of transcription functions of the p53 tumor suppressor by the mdm-2 oncogene. Mol. Med. 1:142-152.
\bibitem{Lia}Liang SH,Clarke MF (1999) A Bipartite Nuclear Localization Signal Is Required for p53 Nuclear Import Regulated by a Carboxyl-terminal Domain. J. Biol. Chem. 274:32699-32703.
\bibitem{Mac} McQuarrie DA (1967) Stochastic approach to chemical kinetics. J. Appl. Probab. 4:413-478.
\bibitem{Rao} Rao CV, Wolf DM, Arkin AP (2002) Control, exploitation and tolerance of intracellular noise. Nature 420:231-237.
\bibitem{Mca} McAdams HH, Arkin A (1997) Stochastic mechanisms in gene expression. Proc. Natl. Acad. Sci. USA 94:814-819.
\bibitem{Bla} Blake WJ, Kaern M, Cantor CR, Collins JJ (2003) Noise in eukaryotic gene expression. Nature 422:633-637.
\bibitem{Pre}Press WH, Teukolsky SA, Vetterling WT , Flannery BP (1992) Numerical Recipe in Fortran. Cambridge University Press.
\bibitem{Gill}Gillespie DT (1997) Exact Stochastic Simulation of Coupled Chemical Reactions. J. Phys. Chem. 1977, 31:2340-2361.
\bibitem{Gill1}Gillespie DT (2000) The chemical langevin equation. J. Phy. Chem. 113:297-306.
\bibitem{Her} Schildt H (2002) The complete reference: Java 2. Tata McGraw Hill.
\bibitem{Kam} Kampen NGV (2007) Stochastic processes in Physics and Chemistry, North Holland, Third Edition.
\bibitem{Ban} Bandt C, Pompe B (2002) Permutation entropy - a complexity measure for time series. Phys. Rev. Lett. 88:174102.
\bibitem{Cao} Cao Y, Tung WW, Gao JJ, Protopopescu VA, Hively LM (2004) Detecting dynamical changes in time series using the permutation entropy. Phys. Rev. 70:046217.
\bibitem{Liu} Liu Z (2004) Measuring the degree of synchronization from time series data. Europhys. Lett. 68:19-25.
\bibitem{Pec}  Pecora LM, Caroll TL (1990) Synchronization in chaotic systems. Phys. Rev. Lett. 64:821.
\bibitem{Ram} Ramaswamy R, Singh RKB, Zhou C, Kurths J (2010) Understanding Complex Systems. Springer, 177-193.
\bibitem{Ros}  Rosenblum MG, Pikovsky AS (2004) Controlling synchronization in an ensemble of globally coupled oscillators. Phys. Rev. Lett. 92:114102.
\bibitem{Ros1} Rosenblum MG, Pikovsky AS, Kurths J (1996) Phase synchronization of chaotic oscillators. Phys. Rev. Lett. 76:1804-1807.




\end{thebibliography}
\end{document}